\documentstyle[12pt,aaspp4]{article}
\begin{document}

\title{Binary Source Parallactic Effect in Gravitational Micro-lensing}

\author{Bohdan Paczy\'nski}
\affil{Princeton University Observatory, Princeton, NJ 08544--1001}
\affil{e-mail: bp@astro.princeton.edu}

\begin{abstract}
The first micro-lensing event discovered towards the Small Magellanic Cloud
by the MACHO collaboration (Alcock et al. 1997b) had a very long time
scale, $ t_0 = 123 $ days.  The EROS collaboration (Palanque-Delabrouille
et al. 1997) discovered a $\sim 2.5\%$ brightness
variation with a period $ P = 5.1 $ days.  The OGLE collaboration
(Udalski et al. 1997) established that the variation persists while
the micro-lensing event is over, and the variable star is the one which
has been micro-lensed, not its blend.

The simplest explanation of the periodic variability is in terms of a
binary star with the orbital period $ P_{orb} = 10.2 $ days, with its
component(s) tidally distorted.  Such objects are known as ellipsoidal
variables.  The binary nature should be verified spectroscopically.

Binary motion of the source introduces a parallactic effect into
micro-lensing light curve, and a few examples are shown.  The effect
is relatively strong if the light center and the mass center of a
binary are well separated, i.e. if the binary has a large photometric
dipole moment.  The diversity of
binary parameters is large, and the corresponding diversity of
photometric effects is also large.  The presence or absence of
the effect may constrain the lens mass and its distance from the source.

\end{abstract}

\keywords{
galaxy: halo --
gravitational lensing
}

\section{Introduction}

The searches for micro-lensing events towards the Magellanic Clouds have
as their hopeful goal the determination of what the dark matter is made of
(Paczy\'nski 1996, and references therein).  While just over a handful
of robust micro-lensing events have been detected towards the LMC by the MACHO
collaboration (Alcock et al. 1997a), the interpretation of the
result has not been agreed upon.  The review of the vigorous discussion
of the subject is beyond the scope of this short paper.

A very intriguing event has been recently discovered
towards the Small Magellanic Cloud
by the MACHO collaboration (Alcock et al. 1997b), and
confirmed by the
EROS collaboration (Palanque-Delabrouille et al. 1997).
It had a very long time scale, $ t_0 = 123 $ days (247 days on the MACHO
scale).  It has been found to be a periodic variable
with the amplitude $ A_{ell} = 0.025 $ and the period of 5.1 days
(Palanque-Delabrouille et al. 1997).
However, it was not clear what varied, the lensed star or its unresolved
blend.  This has been
clarified by the OGLE collaboration (Udalski et al. 1997), which
resolved the two stars thanks to a smaller pixel size and a better
seeing: the variable is the lensed star.
The variability continues with the
same amplitude as it was during the micro-lensing event.

The persistence of periodic variability
implies that the lensed star is likely to be an
ellipsoidal variable (Udalski et al. 1997).  Such stars are common.
Their variability is a result of a tidal distortion
in a binary with the orbital period twice the
photometric period, in this case 10.2 days.

An alternative explanation for the periodic variability might be 
a slowly pulsating B stars (SPB) as discovered by the Hipparcos
mission (Waelkens et al. 1997).  However, such oscillations, just
as the $ \beta $ Cephei pulsations (Dziembowski \& Pamiatnych 1993),
are likely to be driven by iron opacity, and therefore they
are not likely to be present in the SMC.  In any case,
the nature of the periodic oscillations, ellipsoidal variability
or SPB, should be settled with spectroscopic observations.  This
paper is based on the assumption that the lensed star
is a binary ellipsoidal variable.

The purpose of this paper is to describe 
parallactic effect due to binary motion of the source.
This is analogous to the parallactic effect of Earth's orbital
motion (Gould 1992, Alcock et al. 1995).

\section{Binary parallactic effect}

Consider a source which is a binary, with the orbital period $ P_{orb} $,
and a circular orbit with the angular diameter $ \varphi _a $.  The
orbit has an inclination $ i $ to the celestial plane, with $ i = 90^{\circ} $
corresponding to the observer located in the orbital plane.
The orbit appears as
an ellipse with the minor axis $ \varphi _b = \varphi _a \cos i $.
The angle between the major axis and the direction of the source's
proper motion with respect to the lens is $ \alpha $.  An example of
such lensing geometry is shown in Fig.1, where the two binary components
have equal mass, and the two orbital angles are:
$ i = 45^{\circ} $, $ \alpha = 45^{\circ} $.

\begin{figure}[t]
\plotfiddle{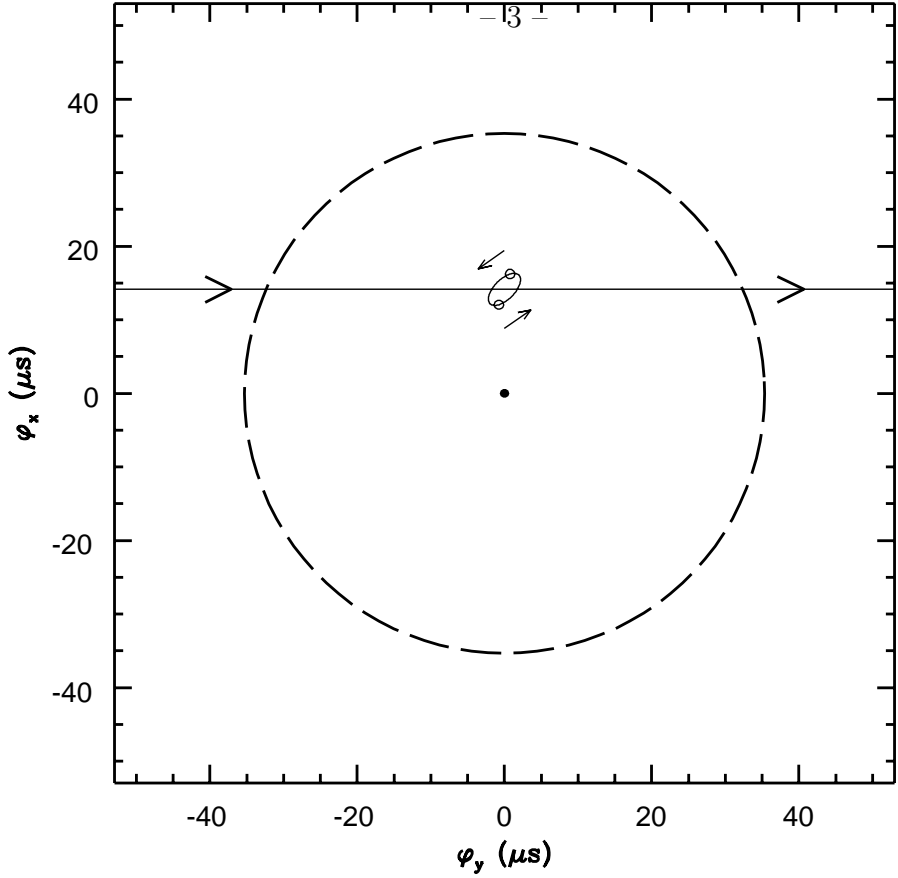}{10cm}{0}{80}{80}{-250}{-160}
\caption{
Geometry of micro-lensing of a binary source by a single point mass,
shown as a black point
located at the center of a dashed circle, its Einstein ring.
The binary source moves along the straight horizontal line with the
impact parameter $ u_{min} = 0.4 $.  The micro-lensing time scale is
$ t_0 = 123 $ days.  The orbital period is $ P_{orb} = 10.2 $ days.
The binary orbit is shown for $ i = 45^{\circ} $ and $ \alpha = 45^{\circ} $.
}
\end{figure}

For the purpose of this analysis I shall adopt a hypothesis that
the lens is in the SMC, and that the relative proper motion of
the source with respect to the lens is $ \dot \varphi \approx
 30 ~ {\rm km ~ s^{-1} ~ / ~ 60 ~ kpc } $.  The angular Einstein ring 
radius is
\begin{equation}
\varphi _E = \dot \varphi t_0 = 36 ~ \mu s \times 
\left( { V \over 30 ~ {\rm km ~ s^{-1}} } \right) ,
\hskip 1.0cm t_0 = 123 ~ {\rm days} .
\end{equation}
Let the two stellar masses be
$ M_1 = M_2 = 2 ~ {\rm M_{\odot}} $, and the orbital
period $ P_{orb} = 10.2 $ days.  The diameter of the circular orbit is
\begin{equation}
A = \left[ G \left( M_1 +M_2 \right) \right] ^{1/3}
\left( { P_{orb} \over 2 \pi } \right) ^{2/3} =
2.2 \times 10^{12} ~ {\rm cm } ~ \left( { M_1 +M_2 \over 4 ~ M_{\odot} } 
\right) ^{1/3} .
\end{equation}
The corresponding angular orbital diameter at the distance of 60 kpc is
\begin{equation}
\varphi _a \equiv { A \over 60 ~ {\rm kpc} } = 2.4 ~ \mu s ~
\left( { M_1 +M_2 \over 4 ~ M_{\odot} } \right) ^{1/3} .
\end{equation}

\begin{figure}[t]
\plotfiddle{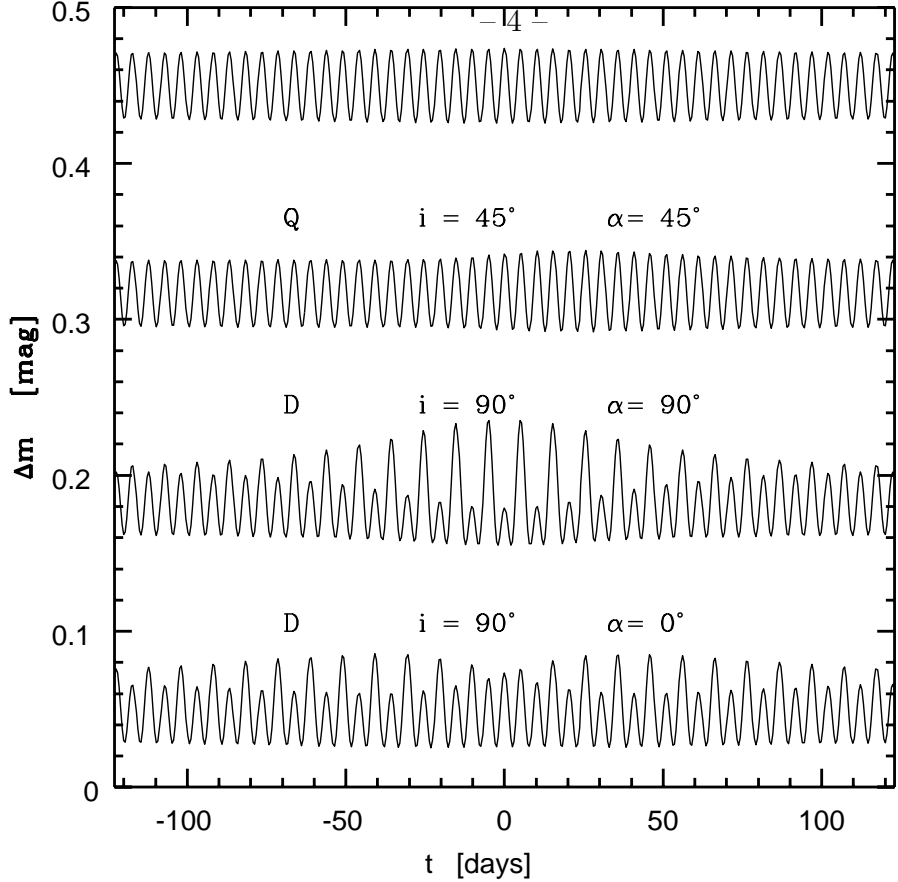}{10cm}{0}{80}{80}{-250}{-160}
\caption{
The differences between four model light curves and a simple
micro-lensing light curve with a blend is shown, with each light curve
shifted by 0.1 mag, for clarity of the display.  D and Q indicate
dipole and quadrupole variations (cf. eqs. 14, 15).
}
\end{figure}

A model light curve was calculated for the recent SMC event
assuming that the blend contributed 26\% to the total light at minimum
(Udalski et al. 1997), the impact parameter was $ u_{min} = 0.4 $,
the lensed source
had its brightness modulated with the period $ P_{orb}/2 = 5.1 $ days, and
the amplitude $ A_{ell} = 0.025 $.
First, no lensing effects due to binary motion of the
source were taken into account.  The difference between the modulated light
curve and a micro-lensing light curve with identical parameters but
without any modulation is shown at the top of Fig.2 and Fig.3.

If the periodic light modulation is due to tidal distortion
of the binary components,
then the periodic displacement of the two stars with respect to the
binary center of mass  modifies the lensing magnification.  Let the
fractional
contribution of the two components and the blend to the minimum light be:
$ f_1, ~ f_2, ~ f_3 $, respectively, with $ f_1 + f_2 + f_3 = 1 $,
and the blend contribution $ f_3 = 0.26 $ (Udalski et al. 1997).
Let the two components have the mass fractions $ m_1 , ~ m_2 $,
respectively, with $ m_1 + m_2 = 1 $, and the total mass $ M_{tot}
= 4 ~ {\rm M_{\odot}} $.  The binary center of mass moves along a straight
line along the `x' direction.  The time dependence of the
two angular coordinates is given as
\begin{equation}
\varphi _{x,cm} = \dot \varphi \left( t - t_{max} \right) , \hskip 1.0cm
\varphi _{y,cm} = \varphi _{y0} = u_{min} \varphi _E ,
\end{equation}
where $ t_{max} $ is the time of maximum magnification.
The primary component moves around the binary center of mass according to
\begin{equation}
\Delta \varphi _{x1} = \Delta \varphi _{a1} \cos \alpha -
\Delta \varphi _{b1} \sin \alpha , 
\hskip 1.0cm
\Delta \varphi _{y1} = \Delta \varphi _{a1} \sin \alpha +
\Delta \varphi _{b1} \cos \alpha ,
\end{equation}
where
\begin{equation}
\Delta \varphi _{a1} = 
m_2 \varphi _a \cos \left( { 2 \pi t \over  P_{orb} } \right) ,
\hskip 1.0cm
\Delta \varphi _{b1} = 
m_2 \varphi _a \sin \left( { 2 \pi t \over P_{orb} } \right) \cos i ,
\end{equation}
and the secondary's orbit is given as
\begin{equation}
\Delta \varphi _{x2} = - { m_1 \over m_2 } \Delta \varphi _{x1} , 
\hskip 1.0cm
\Delta \varphi _{y2} = - { m_1 \over m_2 } \Delta \varphi _{y1} , 
\end{equation}
Finally, the trajectories of both components are given as
\begin{equation}
\varphi _{x,i} = \varphi _{x,cm} + \Delta \varphi _{xi} ,
\hskip 1.0cm
\varphi _{y,i} = \varphi _{y,cm} + \Delta \varphi _{yi} ,
\hskip 1.0cm
i = 1, ~ 2 .
\end{equation}

\begin{figure}[t]
\plotfiddle{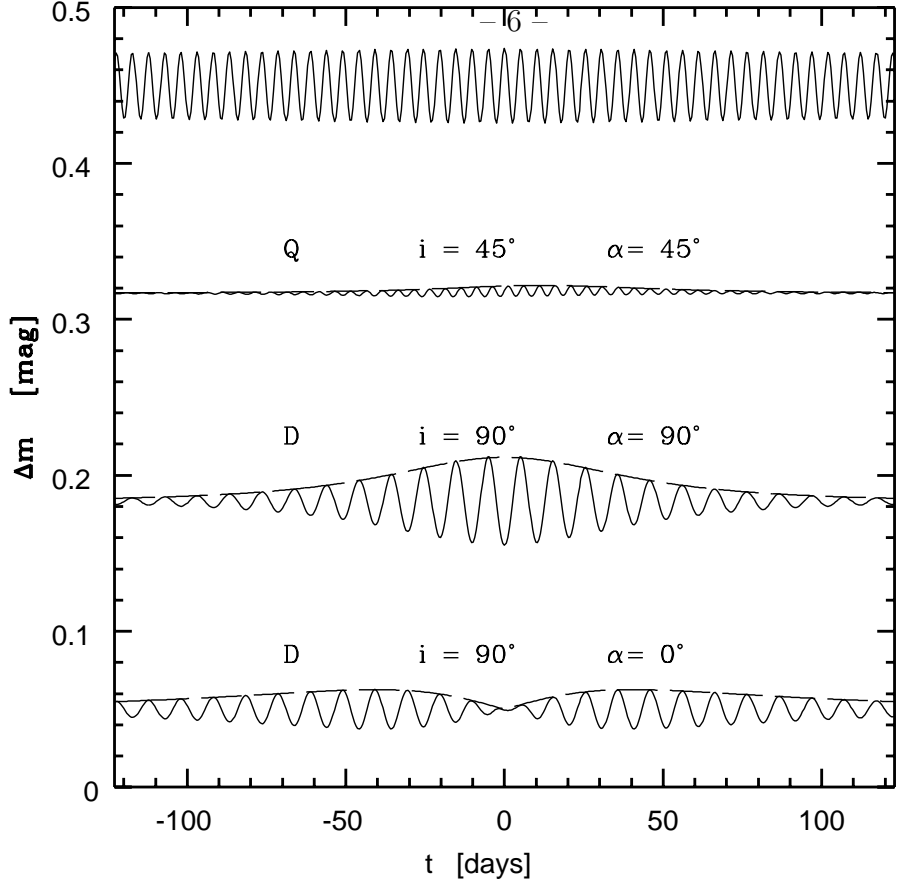}{10cm}{0}{80}{80}{-250}{-160}
\caption{
The uppermost light curve is identical to the uppermost light curve in
Fig.3.  The following 3 light curves are the differences between the
corresponding light curves in Fig.2 and the uppermost light curve.
The dashed lines represent the amplitudes of oscillations calculated
with equations (16).  D and Q stand for dipole and quadrupole variations.
}
\end{figure}

The magnification with respect to the combined non-lensed brightness of the
binary and the blend can be calculated as
\begin{equation}
A = ( A_1 f_1 + A_2 f_2 ) 
\left[ 1 + A_{ell} \cos \left( { \pi t \over P_{orb} } \right) \right] + f_3 ,
\end{equation}
where
\begin{equation}
A_i = { u_i^2 +2 \over u_i \sqrt{ u_i^2 + 4 }} ,
\hskip 1.0cm
u_i^2 = { \varphi _{x,i}^2 + \varphi _{y,i}^2 \over \varphi _E^2 } ,
\hskip 1.0cm
i = 1, ~ 2 .
\end{equation}
Note, that the ellipsoidal light variations with the amplitude $ A_{ell} $
reach the local light maximum at the time of maximum angular separation between
the two components, when the two stars are seen `sideways' and the tidal
distortion makes them appear somewhat larger, and brighter, than a quarter
of the orbital period later (or earlier).

A few examples of binary modulation of a micro-lensing light curve are
are shown in Fig.2 and Fig.3.  In both figures
the uppermost curve corresponds to
the photometric modulation due to 
binary components' tidal distortion.
The following three light curves present the effects
of binary source modulation of the lensing.
In Fig.2 these are the differences between the four models
and the standard micro-lensing light curve.
In Fig.3 the three lower light curves are the differences
between the corresponding curves in Fig.2, and the topmost curve.
The first of these three corresponds to a binary
with both components having identical masses and luminosities, 
and the orbit diameter set to be 0.067 of the Einstein ring radius,
following equations (1) and (3).
The last two curve corresponds to a binary with only one bright component 
separated from the center of mass by 0.013 Einstein ring radii.
All curves are labeled with the values of orbital angles $ i $ and $ \alpha $.

\section{Discussion}

The main practical problem with the binary parallactic effect is the
very large number of parameters which make model fitting impractical
unless the binary nature of the source is constrained with spectroscopic
observations.  In the case of the recent SMC event the spectroscopy is needed
to verify the hypothesis that the source is a binary.

Fig.2 and Fig.3 demonstrate that the effect of binary modulation
is likely to be small in the recent
SMC event.  Therefore, the amplitude of the binary effect can be estimated
expanding the binary modulation in a power series by writing equation (9) as
\begin{equation}
A \approx \left[ A_0 (f_1+f_2) + A' h \left( f_1 m_2 - f_2 m_1 \right) +
0.5 ~ A'' h^2 \left( f_1 m_2^2 + f_2 m_1^2 \right) \right] + f_3 ,
\end{equation}
\begin{equation}
A_0 = { u^2 +2 \over u \sqrt{ u^2 + 4 }} ,
\hskip 1.0cm
A' \equiv { dA \over du } = - { 8 \over u^2 (u^2 +4)^{3/2} } ,
\hskip 1.0cm
A'' \equiv { d^2A \over du^2 } = { 8 ( 5 u^2 + 8 ) \over u^3 (u^2+4)^{5/2} } ,
\end{equation}
where
\begin{equation}
h = { \varphi _a \over \varphi _E } 
\left[ 1.0 - \cos ^2 ( \alpha - \beta ) \sin ^2 i \right] ^{1/2} ,
\hskip 1.0cm
\tan \beta = { \varphi _{y,cm} \over \varphi _{x,cm} } .
\end{equation}
The angle $ \beta $ is between the line joining the lens and the
binary center of mass and the trajectory of that center.
The small range of distances from the lens covered
by the orbital ellipse is calculated as $ h $.  The lens
magnification is expanded in a power series, with the first two
small terms corresponding to the dipole and quadrupole moments of the
binary light distribution:
\begin{equation}
D = h \left( f_1 m_2 - f_2 m_1 \right), \hskip 1.0cm
Q = h^2 \left( f_1 m_2^2 + f_2 m_1^2 \right) ,
\end{equation}
The corresponding amplitudes of dipole and quadrupole variations
are calculated as
\begin{equation}
A_D = A' D , \hskip 1.0cm  A_Q = 0.5 ~ A'' Q .
\end{equation}
The amplitudes calculated with these formulae are shown with dashed lines
in Fig.3.

If the lensed binary has a photometric dipole moment then the
lensing variations are much larger than in the case when there is
no dipole, as quadrupole variations are very small (cf. Fig.3).
It follows that binary lens modulation is much stronger if the two 
components are not identical.
If future spectroscopic observations of the SMC star confirm its
binary nature, then it will be very important to determine not
only the amplitude of the orbital radial velocity variations, but
also to estimate the luminosity ratio of the two components.

Only MACHO collaboration has the full event well covered photometrically.
It will be very interesting to find out if that data will
provide useful information
about the presence or absence of the binary parallactic effect.   The
presence or absence of the effect will provide additional constraint
on the location of the lens.

\acknowledgments{
It is a great pleasure to acknowledge the hospitality of Dr. M. Kubiak,
Director, and all personnel of the Warsaw University Observatory, 
where this paper was written.
This work was supported by the NSF grants AST--9313620 and AST--9530478,
and the Polish KBN grant 2P03D.012.12 to Dr. M. Jaroszy\'nski.}



\end{document}